\documentclass[tikz,11pt,english,a4paper]{article}
\usepackage{float}
\usepackage{jcappub}
\usepackage[utf8]{inputenc}

\usepackage{bm}
\usepackage{algorithm}
\usepackage{nicefrac}
\usepackage{bbold}
\usepackage{dsfont}
\usepackage{setspace}
\usepackage[linewidth=1pt]{mdframed}
\usepackage{bold-extra}
\usepackage{tabto}
\usepackage{amssymb}

\usepackage[noend]{algpseudocode}
\makeatletter
\def\BState{\State\hskip-\ALG@thistlm}
\makeatother

\usepackage{fontawesome5}
\usepackage{hyperref}
\makeatletter
\newcommand{\github}[1]{%
   \href{#1}{\faGithub}%
}
\makeatother

\newcommand{\client}{CLiENT}

\newcommand{\tf}{TensorFlow}

\usepackage{siunitx}
\DeclareSIUnit \parsec {pc}

\usepackage[T1]{fontenc}

\usepackage{lmodern} 
\rmfamily 
\DeclareFontShape{T1}{lmr}{b}{sc}{<->ssub*cmr/bx/sc}{}
\DeclareFontShape{T1}{lmr}{bx}{sc}{<->ssub*cmr/bx/sc}{}

\newcommand{\addfigure}[3]{\begin{figure}[tb]\centering\includegraphics[width=\textwidth]{#1}\caption{\textsl{#3}}\label{#2}\end{figure}}

\usepackage{soul}

\usepackage[edges]{forest}
\definecolor{folderbg}{RGB}{124,166,198}
\definecolor{folderborder}{RGB}{110,144,169}
\newlength\Size
\tikzset{%
  folder/.pic={%
    \filldraw [draw=folderborder, top color=folderbg!50, bottom color=folderbg] (-1.05*\Size,0.2\Size+5pt) rectangle ++(.75*\Size,-0.2\Size-5pt);
    \filldraw [draw=folderborder, top color=folderbg!50, bottom color=folderbg] (-1.15*\Size,-\Size) rectangle (1.15*\Size,\Size);
  },
  file/.pic={%
    \filldraw [draw=folderborder, top color=folderbg!5, bottom color=folderbg!10] (-\Size,.4*\Size+5pt) coordinate (a) |- (\Size,-1.2*\Size) coordinate (b) -- ++(0,1.6*\Size) coordinate (c) -- ++(-5pt,5pt) coordinate (d) -- cycle (d) |- (c) ;
  },
}
\forestset{%
  declare autowrapped toks={pic me}{},
  declare boolean register={pic root},
  pic root=0,
  pic dir tree/.style={%
    for tree={%
      folder,
      font=\ttfamily,
      grow'=0,
    },
    before typesetting nodes={%
      for tree={%
        edge label+/.option={pic me},
      },
      if pic root={
        tikz+={
          \pic at ([xshift=\Size].west) {folder};
        },
        align={l}
      }{},
    },
  },
  pic me set/.code n args=2{%
    \forestset{%
      #1/.style={%
        inner xsep=2\Size,
        pic me={pic {#2}},
      }
    }
  },
  pic me set={directory}{folder},
  pic me set={file}{file},
}

\usepackage{newfloat}
\usepackage{caption}
\DeclareFloatingEnvironment[fileext=frm,placement={!t},name=Algorithm]{pseudoenv}
\DeclareCaptionLabelSeparator{normal-colon}{\normalfont : } 
\newenvironment{pseudo}[2]{
    \gdef\tempcaption{#1}
    \gdef\templabel{#2}
    \begin{pseudoenv}[tb]
    \begin{mdframed}[roundcorner=10pt, middlelinewidth=1pt]
    \begin{center}
    \begin{tabular}{l|l}}
    {\end{tabular}
    \end{center}
    \end{mdframed}
    \caption{\tempcaption}
    \label{\templabel}
    \end{pseudoenv}
    }

\makeatletter
\g@addto@macro\bfseries{\boldmath}
\makeatother

\makeatletter
\def\old@comma{,}
\catcode`\,=13
\def,{%
  \ifmmode%
    \old@comma\discretionary{}{}{}%
  \else%
    \old@comma%
  \fi%
}
\makeatother

\begin{document}


\title{Posterior sampling in the Age of Emulators}

\author[a]{Andreas Nygaard,}
\author[a]{Luca Janken,}
\author[a]{Steen Hannestad,}
\author[a]{and Thomas Tram}

\affiliation[a]{Department of Physics and Astronomy, Aarhus University,
 DK-8000 Aarhus C, Denmark}

\emailAdd{andreas@phys.au.dk}
\emailAdd{andreas.hansen@uzh.ch}
\emailAdd{luca.janken@post.au.dk}
\emailAdd{steen@phys.au.dk}
\emailAdd{thomas.tram@phys.au.dk}

\abstract{We investigate posterior sampling strategies for cosmological parameter inference using fully differentiable neural-network likelihood emulators, which provide both rapid likelihood evaluations and automatic differentiation.

We compare Metropolis–Hastings (MH), the Metropolis-Adjusted Langevin Algorithm (MALA), Hamiltonian Monte Carlo (HMC), the No U-Turn Sampler (NUTS), and Affine Invariant Ensemble Sampling (AIES) using likelihood emulators constructed with the \client{} framework. The methods are tested on emulators of both the $\Lambda$CDM model and a sterile-neutrino extension.

While NUTS generally converges in the fewest samples, its higher computational cost reduces this advantage when performance is measured by wall time. As a result, MALA and even standard MH remain highly competitive. We further find that whitening and covariance adaptation substantially improve sampling efficiency.

The TensorFlow implementations developed for this work are released as the BEST (Batched Emulator Sampling with TensorFlow) package, providing a general framework for sampling arbitrary TensorFlow likelihood functions. The package is available through PyPI as \texttt{best-inference} and on GitHub \github{https://github.com/AndreasNygaard/best-inference.git}.}

\maketitle

\section{Introduction}\label{sec:introduction}
Bayesian parameter inference is a cornerstone of modern cosmology, but its computational cost is often dominated by repeated evaluations of theoretical prediction codes. Since these calculations are typically expensive and rarely provide reliable derivatives of the likelihood, cosmological analyses have historically relied on gradient-free sampling methods such as Metropolis--Hastings~\cite{metropolis,hastings} (MH). Over time, this practical limitation has shaped both software and methodology, making random-walk-based Markov Chain Monte Carlo methods the standard approach for cosmological inference.

The situation is now beginning to change. A growing number of efforts aim to make cosmological inference pipelines differentiable, thereby enabling the use of gradient-based sampling algorithms. Recent examples include differentiable Boltzmann solvers such as ABCMB~\cite{Zhou:2026eit} and broader efforts to construct differentiable inference pipelines. At the same time, emulators have become increasingly common in cosmology, primarily as a means of accelerating expensive theoretical calculations. Most existing emulators target intermediate observables rather than the final likelihood~\cite{Nygaard:2022wri,SpurioMancini:2021ppk,Gunther:2025xrq,Bonici:2023xjk}. While highly successful, such approaches are generally limited by speed of the relevant likelihood codes, and the availability of gradients hinges on the likelihoods being implemented in a compatible framework.

An alternative strategy is to emulate the likelihood directly. This approach has recently been demonstrated with the \client{} framework~\cite{Janken:2025wlq}, which replaces the entire parameter-to-likelihood pipeline with a neural-network surrogate. In addition to providing likelihood evaluations that are orders of magnitude faster than the original pipeline, such emulators are naturally differentiable and can be evaluated efficiently in large batches. Related ideas have also been explored with Gaussian-process likelihood emulators, for example in GPry~\cite{Gammal:2022eob}, although such approaches become increasingly challenging to scale as the dimensionality of the parameter space grows.

The emergence of differentiable likelihood emulators raises a natural question: how much do gradients actually help for posterior sampling? For many years, the lack of gradients has encouraged the view that gradient-based samplers would dramatically outperform traditional methods if only derivatives were available. Indeed, algorithms such as the Metropolis-Adjusted Langevin Algorithm~\cite{mala} (MALA), Hamiltonian Monte Carlo~\cite{hmc} (HMC), and the No U-Turn Sampler~\cite{nuts} (NUTS) are known to generate samples with substantially lower autocorrelation than conventional random-walk samplers. However, they also require additional computations per sample. When likelihood evaluations themselves become extremely inexpensive, the balance between sample quality and sample throughput may be quite different from the regime in which these algorithms were originally developed and analysed.

In this paper, we investigate posterior sampling in this new emulator-driven setting. Using fully differentiable neural-network likelihood emulators constructed with the \client{} framework, we compare MH, MALA, HMC, NUTS, and Affine Invariant Ensemble Sampling~\cite{10.2140/camcos.2010.5.65} (AIES) on both $\Lambda$CDM and sterile-neutrino likelihoods. Rather than focusing solely on the number of likelihood evaluations or effective sample sizes, we assess performance using the practical quantity of greatest interest to end users: the wall-clock time required to obtain converged posterior constraints.

The paper is organised as follows. In section~\ref{sec:sampling} we review the sampling methods considered in this work, and in section~\ref{sec:implementation} we discuss their implementation. Section~\ref{sec:results} describes the numerical implementation, convergence diagnostics, and performance metrics, and presents the comparison of the different samplers. Finally, we summarise our findings in section~\ref{sec:conclusion}.

\section{Sampling methods}\label{sec:sampling}

We employ several different MCMC methods in order to test their capabilities and how well they can exploit the computational benefits of a neural network emulator. Both the auto-differentiation and vectorisation of the emulator enable more efficient methods compared to those used in standard cosmological parameter inference.

\subsection{Gradient-based sampling}\label{sec:gradient}

While traditional MCMC methods have optimal acceptance rates between 0.2 and 0.4~\cite{Gelman1996}, the acceptance rate of gradient-based MCMC methods are typically above 0.65, with some variants even having an acceptance rate close to 1~\cite{2023arXiv231207215R}. The most widely used class of gradient-based samplers are Hamiltonian Monte Carlo (HMC) samplers which explores the parameter space using Hamiltonian dynamsinics. This is achieved by treating the likelihood landscape as an energy landscape and sampling points along the trajectories of particles moving according to the equations of motion defined by the Hamiltonian of the system:
\begin{equation}\label{eq:hamiltonian}
	\mathcal{H}(\bm{x}, \bm{p}) = U(\bm{x}) + \frac{1}{2}\bm{p}^\top M^{-1} \bm{p}\;,
\end{equation}
where the first term, $U(\bm{x}) = -\log{\left(\pi(\bm{x})\right)}$, is the potential energy (the energy landscape) related to the probability density, $\pi(\bm{x})$, while the second term is the kinetic energy of the particles with a symmetric and positive definite mass matrix, $M$. Each step proposes a long trajectory based on the gradients and the momentum variables sampled from the mass matrix, which makes the sampling very efficient since it explores regions of high probability in an almost deterministic way. The only stochasticity in the sampling comes from resampling of the momentum variables and numerical errors in the leap-frog integrator.

There are numerous variants of HMC and one that is often used is the No U-Turn Sampler (NUTS), which automatically controls the number of leap-frog steps. It does so, as the name implies, by making sure the trajectories do not perform a u-turn. A binary tree is built for both forward and backwards moving trajectories and the leap-frog integrator is halted as soon as the u-turn criterion is fulfilled, i.e., the momentum of either the forward or backwards moving particle begins to point to the other particle. The next MCMC point is then sampled uniformly from all valid points (preserving detailed balance) in the trajectory in the binary tree, and the process repeats from that new point. 

Besides variants of HMC, there are various other types of gradient-based MCMC methods to choose from. A different approach is the Metropolis-Adjusted Langevin Algorithm (MALA) utilising Langevin dynamics for sampling. The proposal can be derived from the Langevin equation,
\begin{equation}
	m \frac{{\rm d}^2 \bm{x}}{{\rm d} t^2} = -\lambda \frac{{\rm d}\bm{x}}{{\rm d}t} - \nabla U(\bm{x}) + \eta(t)\; ,
\end{equation} 
where the $\bm{x}$ is the position of a particle with mass $m$, $\lambda$ is the damping coefficient, $U(\bm{x})$ is the potential, and $\eta(t)$ is a Gaussian white-noise process. MALA corresponds to the overdamped limit, $m\rightarrow0$, where inertia disappears and we are left with Brownian motion and drift due to the gradients. The potential is the same as for HMC, $U(\bm{x}) = -\log{\left(\pi(\bm{x})\right)}$, and the noise term can be rescaled to a multivariate normal distribution. Discretising this differential equation thus yields the proposal
\begin{equation}
	\bm{x}_{n+1} = \bm{x}_n + \epsilon\nabla \log{\left(\pi(\bm{x}_n)\right)} + \sqrt{2\epsilon} \xi_n\;,
\end{equation}
where $\xi_n \sim \mathcal{N}(0,\mathbb{1})$, and $\epsilon$ is the step size. This proposal corresponds to a normal distribution with mean $\bm{\mu} = \bm{x}_n + \epsilon\nabla \log{\left(\pi(\bm{x}_n)\right)}$ and covariance $\Sigma = 2\epsilon\mathbb{1}$, thus shifting the proposal distribution in the direction of the gradients compared to the standard Metropolis--Hastings proposal.

\subsection{Mass matrix or canonical transformation}\label{sec:mass_matrix}

Gradient-based MCMC methods that are variants of HMC typically all require a mass matrix, $M$, as shown in equation~\eqref{eq:hamiltonian} in order to work efficiently. In the \tf{} implementation of HMC\footnote{Defined in the TensorFlow-Probability package.}, however, there is no option for providing a mass matrix to the sampler. This means that the mass matrix of the system is essentially the identity, and the momentum components along each parameter direction are thus similar in magnitude since the momentum is drawn from $\mathcal{N}(\bm{0}, M)$. This is not ideal when parameters have different scales due to the step size being dominated by the parameters with the smallest scales, leading to very suboptimal sampling in parameters of larger scales. In order to accommodate this, we can transform the parameters to a whitened parameter space using the Cholesky decomposition of the posterior covariance matrix, $\Sigma$~\cite{Radford2011}:
\begin{align}
	\Sigma &= LL^\top\;, \\
	\bm{x} &= \bm{\mu} + L\bm{z}\;,
\end{align} 
where $L$ is the lower triangular matrix obtained from the Cholesky decomposition, $\bm{\mu}$ is the posterior mean, which we want to subtract from the position coordinate to center the distribution around the origin, and $\bm{z}$ is the transformed position vector in the whitened parameter space. In order to ensure the properties of the Hamiltonian system, we want the transformation to be canonical~\cite{2024arXiv240307495T}. This requires preserving the canonical symplectic 2-form, $\omega = {\rm d}x \bigwedge {\rm d}p$, with the corresponding transformation of the momentum space:
\begin{equation}
	\bm{p} = \left(L^\top\right)^{-1}\bm{q}\;,
\end{equation}
where $\bm{q}$ is the transformed momentum vector. These transformations ensure the preservation of the phase-space volume~\cite{arnold1989mathematical,Goldstein2002}, and since the mass matrix is roughly the inverse of the posterior covariance, i.e., $M = \Sigma^{-1}$~\cite{2017arXiv170102434B}, the transformations also result in the mass matrix in the whitened space being exactly the identity:
\begin{equation}
	\bm{p}^\top M^{-1} \bm{p} = \bm{q}^\top L^{-1} \Sigma \left(L^\top\right)^{-1}\bm{q} = \bm{q}^\top L^{-1} L L^\top \left(L^\top\right)^{-1}\bm{q} = \bm{q}^\top \mathbb{1} \bm{q}\;.
\end{equation} 
The Hamiltonian in the whitened space is then
\begin{equation}
	\tilde{\mathcal{H}}(\bm{z}, \bm{q}) = \tilde{U}(\bm{z}) + \frac{1}{2}\bm{q}^\top \mathbb{1} \bm{q}\;,
\end{equation}
with $\tilde{U}(\bm{z}) = U(\bm{\mu} + L\bm{z})$ being the potential energy in the whitened space.

We now need to modify the probability function used in the sampling as well. The probability mass is invariant under reparameterisation, which means that
\begin{equation}\label{eq:prob_mass}
	\int_{A} \pi_x(\bm{x}) {\rm d}\bm{x}= \int_{T^{-1}(A)} \pi_z(\bm{z}) {\rm d}\bm{z}\;,
\end{equation}
where $T: \mathbb{R}^n \rightarrow \mathbb{R}^n$ is the affine transformation defined by $T(\bm{z})=\bm{\mu} + L \bm{z}$, $\pi_x(\bm{x})$ is the probability density in the correlated parameter space, and $\pi_z(\bm{z})$ is the same for the whitened parameter space. We can transform the RHS of equation~\eqref{eq:prob_mass} to the whitened space using the determinant of the Jacobian of the transformation~\cite{apostol_calculus_vol2}, i.e., $|{\rm det}(J_T)| = |{\rm det}(L)|$,
\begin{equation}\label{eq:transform_int}
	\int_{A} \pi_x(\bm{x}) {\rm d}\bm{x} = \int_{T^{-1}(A)} \pi_x\left(T(\bm{z})\right) \left|{\rm det}(L)\right| {\rm d}\bm{x}\;.
\end{equation}
Comparing the RHSs of equations~\eqref{eq:prob_mass} and~\eqref{eq:transform_int}, shows that the probability density, $\pi_z$, in the whitened space is related to the probability density, $\pi_x$, in the correlated space as
\begin{equation}
	\pi_z(\bm{z}) = \pi_x\left(T(\bm{z})\right) \left|{\rm det}(L)\right|\;.
\end{equation}
The logarithm of the probability density (which is what we actually want to sample from) is then
\begin{equation}\label{eq:log_prob_transform}
	\log{\left(\pi_z(\bm{z})\right)} = \log{\left(\pi_x\left(T(\bm{z})\right)\right)} + \log{\left(\left|{\rm det}(L)\right|\right)}\;,
\end{equation}
where the last term is just a constant defined by the transformation. Omitting the constant does not change the sampling procedure, but we would then need to account for it when transforming back to the correlated parameter space. By including it in the function we are sampling from, we ensure that the resulting distribution transforms neatly back to the correlated space using only the transformation $T$. Sampling in a whitened parameter space can be beneficial in other types of sampling besides HMC variants as well, simply because very pronounced covariances in high dimensionality are difficult to sample, and we will use this strategy for all our samplings\footnote{We will not use whitening for the Affine Invariant Ensemble Sampler, since whitening is an affine transformation.}. The procedure for sampling in the whitened space is summed up in Algorithm~\ref{alg:whitened}.

\begin{pseudo}{\textsl{Pseudo-code sampling in whitened space. The probability density function is transformed using equation~\eqref{eq:log_prob_transform}, and the resulting probability density function in the whitened space is used for the Hamiltonian Monte Carlo sampling.}}{alg:whitened}
	\hspace{-1.4em}
	\begin{minipage}{0.69\textwidth}
		\vspace{1.7em}
		\begin{spacing}{1.2}
			\begin{algorithmic}
				\State $\Sigma$\tabto{5em}$=$\hspace{0.5em} posterior covariance matrix
				\State $\mu$\tabto{5em}$=$\hspace{0.5em} posterior mean vector
				\State \texttt{log\_prob}\tabto{5em}$=$\hspace{0.5em} logarithm of probability density function
				\State $L$\tabto{5em}$=$\hspace{0.5em} \texttt{CholeskyDecomposition($\Sigma$)}
				\vspace{1.5em}
				\State \textbf{def} \texttt{log\_prob\_whitened($z$)}:
					\State\tabto{1.5em} $x$\tabto{5.8em}$=$\hspace{0.5em} $\mu \;+\; $\texttt{MatrixMultiplication($L$,$z$)}
					\State\tabto{1.5em} \texttt{log\_det}\tabto{5.8em}$=$\hspace{0.5em} \texttt{sum(log(Diagonal($L$)))}
					\State\tabto{1.5em} \textbf{return} \texttt{log\_prob($x$)}$\;+\;$\texttt{log\_det} 
				\vspace{1.5em}
				\State chain \tabto{5em}$=$\hspace{0.5em} \texttt{Do\_HMC($\pi$=log\_prob\_whitened,
				\State \tabto{9.8em} start\_position=[0,0,...,0])}
				\State chain \tabto{5em}$=$\hspace{0.5em} $\mu\;+\;$\texttt{MatrixMultiplication($L$,}chain\texttt{)}
			\end{algorithmic}
		\end{spacing}
	\end{minipage}
	&
	\hspace{0.07em}
	\begin{minipage}{0.28\textwidth}
		\vspace{8.9em}
		\begin{spacing}{1.2}
			\textit{transformed probability\\ density}
			\vspace{4.5em}
			
			\textit{sampling with HMC\\ 
			\\
			transform back}
		\end{spacing}
	\end{minipage}
\end{pseudo}

\subsection{Ensemble sampling}\label{sec:ensemble}
Another great attribute of neural networks is the very efficient parallelisation through vectorised calls to the networks. As shown in Figure~\ref{fig:scaling}, evaluating the emulator for up to 100 points simultaneously does not affect the evaluation time significantly. We can thus utilise ensemble sampling methods very efficiently.

\addfigure{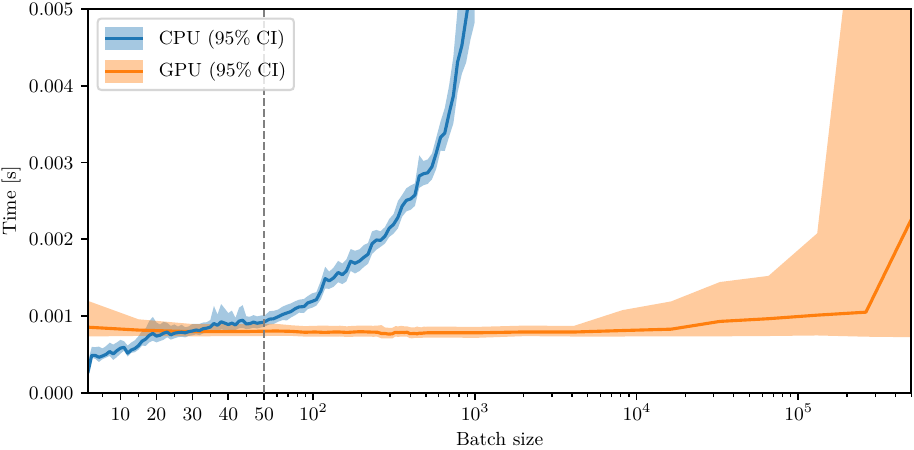}{fig:scaling}{Mean evaluation time of the neural network emulator as a function of batch size on both CPU (14-core Apple M4 Pro) and GPU (NVIDIA Tesla V100). On the CPU, evaluation time increases only slightly for batch sizes of up to 100, which allows for parallel calls with almost no extra costs for low batch sizes. On the GPU, however, the evaluation time remains constant for batch sizes as high as $10^4$, thus making vectorised parallelisation much more efficient. The vertical dashed line marks the transition from a linear axis to a logarithmic axis on the plot.}

There are various sampling methods that can exploit the favourable scaling of vectorised calls to the neural network. Simply using vectorisation to run multiple independent chains in parallel without much extra cost is certainly a possibility, but for standard MCMC methods, such as Metropolis-Hastings, having a large number of chains is not usually beneficial due to the high autocorrelation of the chains, i.e., a larger number of chains do not necessarily lead to faster convergence. For HMC, one can absolutely use multiple chains as well, even though it and its variants are already quite efficient with only a single chain. Running multiple chains is rarely needed, but vectorisation allows for many simultaneous chains at essentially no extra cost, which makes the sampling much faster. The MALA implementation in \tf{} is, likewise, very optimised for multiple chains. For this method, running a large number of chains (50-100) results in much faster convergence of the posterior distribution, so the chains do not need to be as long as those of some of the other methods.

Even though they can utilise vectorised calls to the emulator, these methods are not ensemble methods. The individual walkers or particles simulated in ensemble methods are generally not independent since they exchange information in order to update their position. A widely used class of ensemble sampling methods are the Affine Invariant Ensemble Samplers (AIES), which rely on a large number of walkers exploring the parameter space simultaneously and exchanging information. As the name implies, the sampling process of these methods is invariant to affine transformation such as a Cholesky transformation to whitened space. This means that they perform well without being given any prior information about the correlations and length scales of the parameters. The standard AIES method exchanges information between pairs of walkers and update their position with a so-called \emph{stretch move}~\cite{10.2140/camcos.2010.5.65}:
\begin{equation}
	\bm{x}^{i}_{n+1} = \bm{x}^{j}_n + z(\bm{x}^{i}_{n} - \bm{x}^{j}_{n})\;,
\end{equation}
where $i$ and $j$ refer to a pair of walkers and $z$ is drawn from a distribution with density
\begin{equation}
	g(z) \propto \begin{cases} 		\dfrac{1}{\sqrt{z}}        & \text{if } x\in \left[\dfrac{1}{a}, a\right], \\
                 			        0                               & {\rm otherwise}.    %
      		  \end{cases}
\end{equation}
The parameter $a>1$ is a hyperparameter that is usually set to $a=2$~\cite{2013PASP..125..306F}. In practice, this means drawing a number $u$ from a uniform distribution between 0 and 1, and transforming it by equating it with the cumulative distribution function of $g(z)$. The correctly distributed value of $z$ is thus
\begin{equation}
	z = \frac{1}{a} \left( u(a-1)+1 \right)^2\;.
\end{equation}

Other variants of AIES differ in their \emph{moves}, i.e., how the positions of the walkers are updated. Different moves can be well suited for different scenarios depending on the dimensionality and geometry of the probability surface. The Differential Evolution (DE) move draws inspiration from genetic algorithms and uses information from two walkers to update the position of a third walker. This move takes the form
\begin{equation}
	\bm{x}^{i}_{n+1} = \bm{x}^{i}_n + \gamma(\bm{x}^{j}_{n} - \bm{x}^{k}_{n})\;,
\end{equation}
where $i$, $j$, and $k$ refer to the three walkers used for the move and $\gamma = (1+\eta)2.38/\sqrt{2n_{\rm dim}}$. $\eta\sim\mathcal{N}(0,\sigma_{\gamma})$ is normally distributed with $\sigma_{\gamma}$ as the standard deviation controlling the randomness of the move.

\section{Sampler implementation}\label{sec:implementation}

To complement the \tf{}-based emulator framework, we provide a unified sampling library, BEST (\textbf{B}atched \textbf{E}mulator \textbf{S}ampling with \textbf{T}ensorFlow), which implements a set of scalable MCMC methods designed for GPU execution and emulator-based likelihoods. The package is built to interface directly with any differentiable \tf{} likelihood function and is distributed as a standalone Python package installable via \texttt{pip install best-inference} and also available on GitHub\footnote{Available at: \url{https://github.com/AndreasNygaard/best-inference.git}}. While originally developed in conjunction with the \client{} framework, BEST is fully agnostic to emulator architecture and can be used with any TensorFlow-compatible log-likelihood.

\subsection{Unified sampling interface}

All samplers are exposed through a single high-level interface implemented in the \texttt{Sampler} class, which manages initialisation of walkers, boundary handling, and covariance adaptation. The sampler is initialised as
\begin{verbatim}
sampler = best.Sampler(log_prob_fn, bounds=(lower, upper))
\end{verbatim}
where \texttt{log\_prob\_fn} defines the target log-probability and \texttt{bounds} optionally specify a hyper-rectangular support of the parameter space. When bounds are provided, the likelihood is automatically extended with a smooth exponential penalty outside the domain, ensuring that walkers are discouraged from leaving the physically relevant region while preserving differentiability.

Sampling is performed through a unified call,
\begin{verbatim}
results = sampler.sample(method="hmc", n_steps=..., n_chains=...)
\end{verbatim}
where \texttt{method} selects between MH, AIES, HMC, NUTS, and MALA. The interface further supports flexible burn-in schedules, multiple initialisation strategies of the walker positions (uniform, repeated, or Gaussian), and optional covariance matrix updates during burn-in.

The returned object contains both samples and diagnostics:
\begin{verbatim}
results.samples
results.acceptance_rate
results.evaluations
\end{verbatim}
with optional fields for burn-in trajectories and covariance estimates when adaptive phases are enabled.

\subsection{Covariance adaptation and parameter space whitening}

For MH, HMC, NUTS, and MALA, BEST performs automatic covariance estimation during burn-in. After each burn-in iteration, an empirical covariance matrix is computed from the ensemble of walkers and used to construct a whitening transformation (as in Algorithm~\ref{alg:whitened}) of the parameter space. This improves isotropy of the target distribution and reduces autocorrelation times in highly anisotropic posteriors.

To ensure numerical stability, the covariance estimate is regularised via Cholesky factorisation, enforcing positive definiteness before reuse. The affine-invariant ensemble sampler is excluded from this procedure due to its invariance under affine transformations.

Initial walker configurations can be drawn from either uniform distributions over the parameter bounds, replicated initial states, or Gaussian perturbations around a central point. The Gaussian initialisation is implemented using a hypersphere-based sampler, ensuring consistent scaling across dimensions.

\subsection{Computational implementation and performance}

The sampling kernels are implemented in \tf{} and \tf{}-Probability, with modifications enabling full compatibility with graph execution and XLA just-in-time (JIT) compilation. While \tf{}-Probability provides native implementations of MH, HMC, NUTS, and MALA, their default Python control flow and loop structure prevents efficient compilation. To remedy this, we wrap these kernels in a custom JIT-compatible driver that enables execution entirely within the TensorFlow computation graph. 

Although publicly available implementations of AIES \tf{} do exist~\cite{SpurioMancini:2021ppk}, these are most often not optimised for graph execution like the ones implemented in the \tf{}-Probability package. We therefore provide a custom implementation of AIES, based on the stretch-move update rule, designed specifically for vectorised GPU execution. This implementation achieves approximately $10^4$ steps per second (for all walkers combined) on an NVIDIA Tesla V100 under JIT compilation.

The custom driver allows for very fast sampling with all of the methods with speeds ranging from $\sim\!10^2$ steps per second for NUTS to $\sim\!10^4$ steps per second for AIES and MH. NUTS is the most advanced algorithm and it also utilises a large binary tree of leap-frog steps (requiring gradient computations), so this of course samples slower than gradient-free algorithms.

\begin{table}[]
\centering
\begin{tabular}{r c c c c}
\hline
\hline
     & \: \textbf{CPU + GE} \: & \: \textbf{CPU + JIT} \: & \: \textbf{GPU + GE} \:  & \: \textbf{GPU + JIT} \: \\
\textbf{MH}      &  608 (153)         & 1,287 (207)    & 864 (973)                      & 8,688 (2,932)               \\
\textbf{MALA}  &  131 (29)           & 344 (84)   & 277 (286)                      & 3,534 (1,028)                \\
\textbf{HMC}    &  76 (7)               & 167 (22)   & 123 (129)                       & 1,856 (294)                  \\
\textbf{NUTS}  &  8 (0.8)               & 36 (3.7)  & 13 (9)                            & 153 (46)                         \\
\textbf{AIES}    &  836 (168)          & 1,030 (188)   & 806 (947)                      & 8,238 (2,901)                \\
\hline
\end{tabular}
\caption{\textsl{Sampling speeds (steps per second) across different hardware and execution modes. CPU corresponds to a 14-core Apple M4 Pro, and GPU corresponds to an NVIDIA Tesla V100. GE denotes standard TensorFlow graph execution, while JIT refers to XLA-compiled execution. Values are reported for 20 (1,000) chains.}}
\label{tab:speeds}
\end{table}

\addfigure{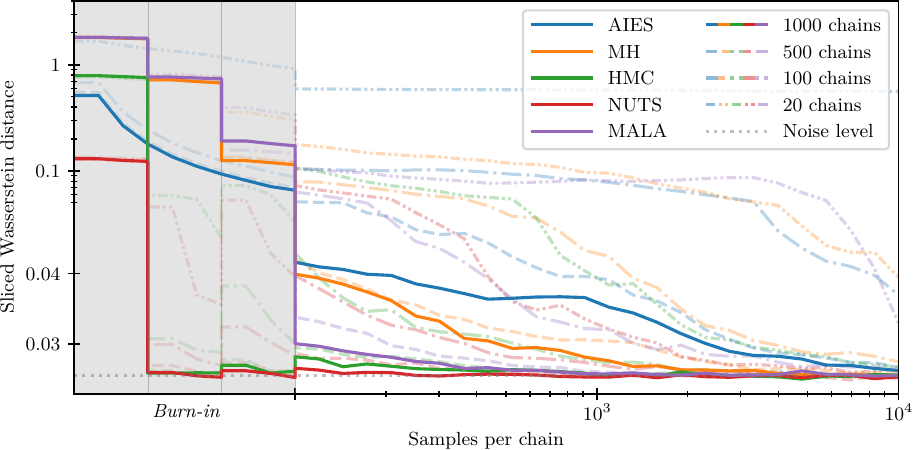}{fig:n_chains}{Sliced Wasserstein distance as a function of samples per chain for each MCMC method and different numbers of chains when sampling with the $\Lambda$CDM emulator. The grey region represents burn-in consisting of three iterations, except for AIES which uses a single extended burn-in phase.}

The speeds of the different samplers are reported in Table~\ref{tab:speeds} for different hardware and compilation settings. Using the GPU with standard TensorFlow graph execution but without JIT compilation yields similar speeds for 20 chains and 1,000 chains, as suggested by Figure~\ref{fig:scaling}. However, enabling JIT compilation introduces a clearer dependence on batch size, resulting in lower throughput for a smaller number of chains. This is likely because XLA generates fused kernels that more efficiently utilise the GPU compute resources even for smaller workloads, whereas standard graph execution relies more heavily on large batch dimensions to achieve high GPU occupancy. As a result, JIT compilation improves overall sampling performance on the GPU, while also making the scaling behaviour with respect to the number of chains more apparent. This demonstrates that large-scale vectorisation on GPUs provides a practical speedup in terms of effective sampling efficiency, even when wall-clock scaling saturates.

\section{Comparison of sampling methods}\label{sec:results}
For this paper, we will compare the performances of five different MCMC sampling methods:
\begin{itemize}
	\item Metropolis-Hastings (MH)
	\item Metropolis-Adjusted Langevin Algorithm (MALA)
	\item Hamiltonian Monte Carlo (HMC)
	\item No U-Turn Sampler (NUTS)
	\item Affine Invariant Ensemble Sampling (AIES) 
\end{itemize}
Each method is tested on both the $\Lambda$CDM model as well as a sterile neutrino extension with two additional parameters, $N_{{\rm eff},s}$ and $m_s$. A likelihood emulator for both models was created using the CLiENT framework~\cite{Janken:2025wlq} and the following data sets:
\begin{itemize}
	\item Planck 2018 high-$\ell$ TTTEEE, low-$\ell$ TT+EE, and lensing~\cite{Planck:2018vyg,Planck:2019nip}.
	\item Baryon Acoustic Oscillations (BAO) measurements from BOSS DR12~\cite{boss2016}, the main galaxy sample of BOSS DR7~\cite{ross2014} and 6dFGS~\cite{Beutler:2011hx}.
\end{itemize}
This results in a 27-dimensional parameter space (including nuisance parameters) for the $\Lambda$CDM model and a 29-dimensional space for the sterile neutrino extension model.

As mentioned in section~\ref{sec:mass_matrix}, some of the MCMC methods require a decent covariance matrix for more optimal sampling, so all samplings are initialised with a burn-in loop where the covariance matrix and the best-fit point are iteratively updated based on short chains. The initial covariance matrix is diagonal with scales proportional to the dimensions of the bounding box of the emulator. AIES does not need a covariance matrix, although one can be used to initialise the walkers more optimally. However, to not impose additional information, we instead initialise the walkers uniformly in the bounding box. We also do not use a burn-in loop for AIES, but rather a single burn-in run of significant length, where the last positions of the walkers are used as initial positions in the final run. This effectively corresponds to a higher number of steps where the first part is discarded. 

\subsection{Convergence metrics}
To test the convergence of the MCMC methods, we use three different metrics; the Gelman-Rubin type convergence statistic~\cite{10.1214/ss/1177011136}, the credible metric introduced in Ref.~\cite{Janken:2025wlq}, and the sliced Wasserstein distance~\cite{2019arXiv190200434K}.

\textbf{ The Gelman-Rubin convergence statistic}, $R$, measures the discrepancy between the variance of chain means and the typical within-chain variance. When all chains are perfectly mixed, this statistic should approach 1 from above, and using $R-1$ instead yields a metric that tends toward zero when the chains reach a fully mixed steady-state. In the multivariate case, the metric can be computed from the covariance, $B$, of the chain means, and the mean, $W$, of chain covariances. Inspired by the implementation in Cobaya~\cite{cobaya}, we can eliminate scale dependence, by scaling these matrices using the diagonal $B$, i.e., $d_i = \sqrt{B_{ii}}$:
\begin{gather}
	\nonumber D = {\rm diag}(d), \\
	                  \tilde{B}=D^{-1}B D^{-1}, \\
	\nonumber \tilde{W}=D^{-1}W D^{-1}.
\end{gather}
The metric is then the maximum absolute eigenvalue of the matrix $\tilde{W}^{-1}\tilde{B}$:
\begin{gather}
	R - 1 = \max_i |\lambda_i|, \; \textrm{for } \lambda_i \in {\rm eig}\!\left(\tilde{W}^{-1} \tilde{B} \right).
\end{gather}
This metric is close to the multivariate potential scale reduction factor (MPSRF) introduced in Ref.~\cite{brooks_gelman}, except for the diagonal normalisation. This metric, however, only measures how well the MCMC sampling has converged to a steady state. It does not measure how similar the resulting posterior is to the ``ground truth''. For this, we need the other metrics.

\textbf{The credible metric} for the $i^{\rm th}$ parameter $\theta_i$ and credible level $\alpha$ is defined as
\begin{equation}
	\Delta_i^{{\rm CM}_{\alpha}} = \frac{|\theta^{\alpha+}_{i} - \theta^{\alpha+}_{i,0}| + |\theta^{\alpha-}_{i} - \theta^{\alpha-}_{i,0}|}{\theta^{\alpha+}_{i,0} - \theta^{\alpha-}_{i,0}},
\end{equation}
where $+$ ($-$) refers to upper (lower) credible limits and the subscript 0 refers to the true credible levels. The metric we will be using is the maximum value of the credible metric across all parameters, i.e. $\Delta^{{\rm CM}_{\alpha}} = \max_i(\Delta_i^{{\rm CM}_{\alpha}})$ for $\alpha=\{68\%, 95\%\}$. The credible metric thus measures how similar the obtained credible limits are to the true limits, but it has a few drawbacks. One of them is that the metric is scale dependent, but this can be alleviated by whitening the chains prior to the computation (the true limits should of course be whitened using the same covariance). Another problem is that the denominator of the metric, i.e., the width of the credible interval, is dependent on the credible level $\alpha$. If the relative error is the same for two credible levels, a 95\% level will thus result in a smaller value of the metric than one would obtain with a 68\% level. This cannot as easily be circumvented so we effectively treat different credible levels as separate metrics. The credible metric is also only sensitive to information from the marginalised 1D posteriors, so possible differences in correlations would not be probed by this.

\textbf{The Wasserstein distance}, also known as the optimal transport distance, measures the minimum cost required to transport probability mass from one distribution in order to transform it into another. As such, it captures the full multidimensional geometry and correlations of the posterior distribution. The exact computation of the d-dimensional Wasserstein distance involves constructing a pairwise cost matrix and solving an optimal transport problem, typically formulated as a linear program. While constructing the cost matrix scales as $\mathcal{O}(dnm)$, where $n$ and $m$ are the number of samples in the two distributions being compared, solving the transport problem itself scales substantially worse and quickly becomes computationally infeasible for more than a few thousand samples due to both runtime and memory requirements. Because of this limitation, we instead employ the \emph{sliced Wasserstein distance}, which averages 1-dimensional Wasserstein distances over many random projections through parameter space. The 1-dimensional Wasserstein distance is significantly cheaper to compute since, for empirical distributions, it reduces to sorting the projected samples and computing their mean absolute difference, avoiding the need for linear programming entirely. Although approximate, the sliced Wasserstein distance still captures multidimensional structure and correlations provided that a sufficiently large number of projections is used. The same set of 500 projection directions is reused for all computations to ensure that equivalent features and correlations of the posterior are probed consistently across all samplers.

The latter two metrics require a comparison with a ``ground truth'' which in our case will be a very long Metropolis-Hastings sampling of 200,000 steps and 100 walkers, where we only keep the last 100,000 points for the comparisons. We use the emulator to obtain this reference run since its accuracy has already been established in Ref.~\cite{Janken:2025wlq}. The stochasticity of the sliced Wasserstein distance and the fact that a finite MCMC sampling will be used as the ``ground truth'' make both the sliced Wasserstein distance and the credible metric approach a noise floor instead of exactly zero. These noise floors can be estimated by computing the metrics for an MCMC sampling similar to the reference run (same number of walkers and steps). This should yield metric values that are as good as we can realistically obtain.

\subsection{$\Lambda$CDM results}

\addfigure{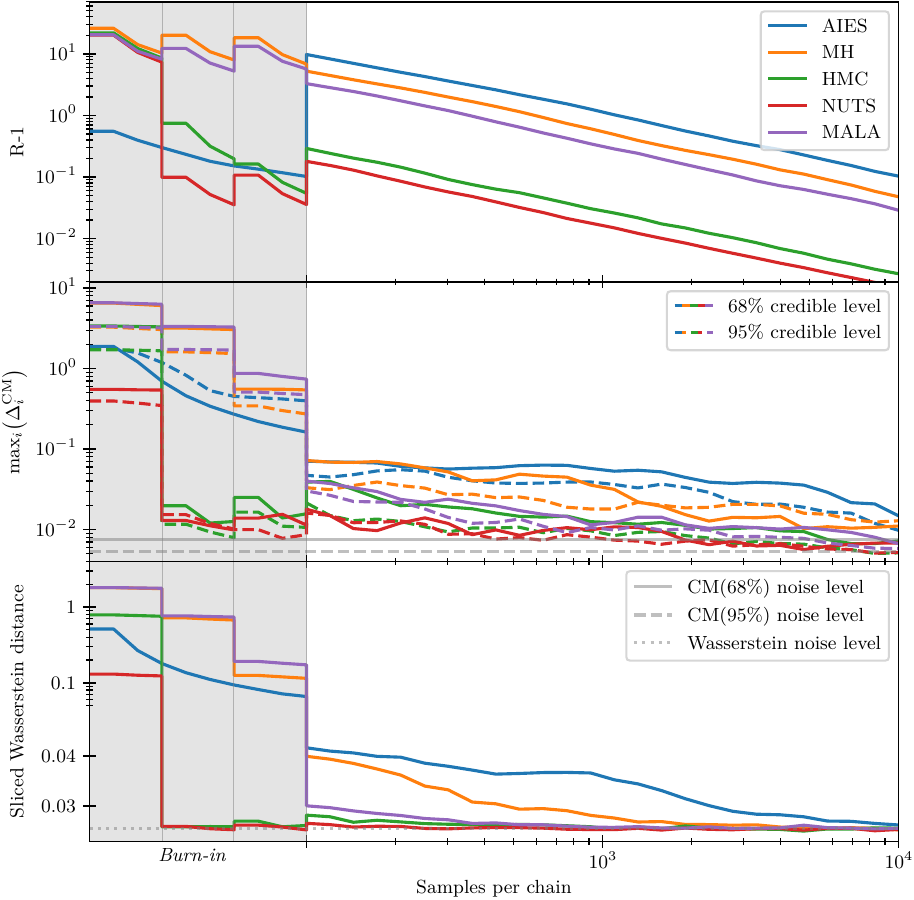}{fig:samples_lcdm}{All metrics as a function of samples per chain for every MCMC method using the $\Lambda$CDM emulator and 1,000 chains. The grey region refers to the burn-in period consisting of three updates of the covariance matrix before the final sampling, except for AIES which only has a single longer burn-in iteration. Noise levels obtained from independent reference runs are also depicted for the credible metric and the sliced Wasserstein distance.}

\addfigure{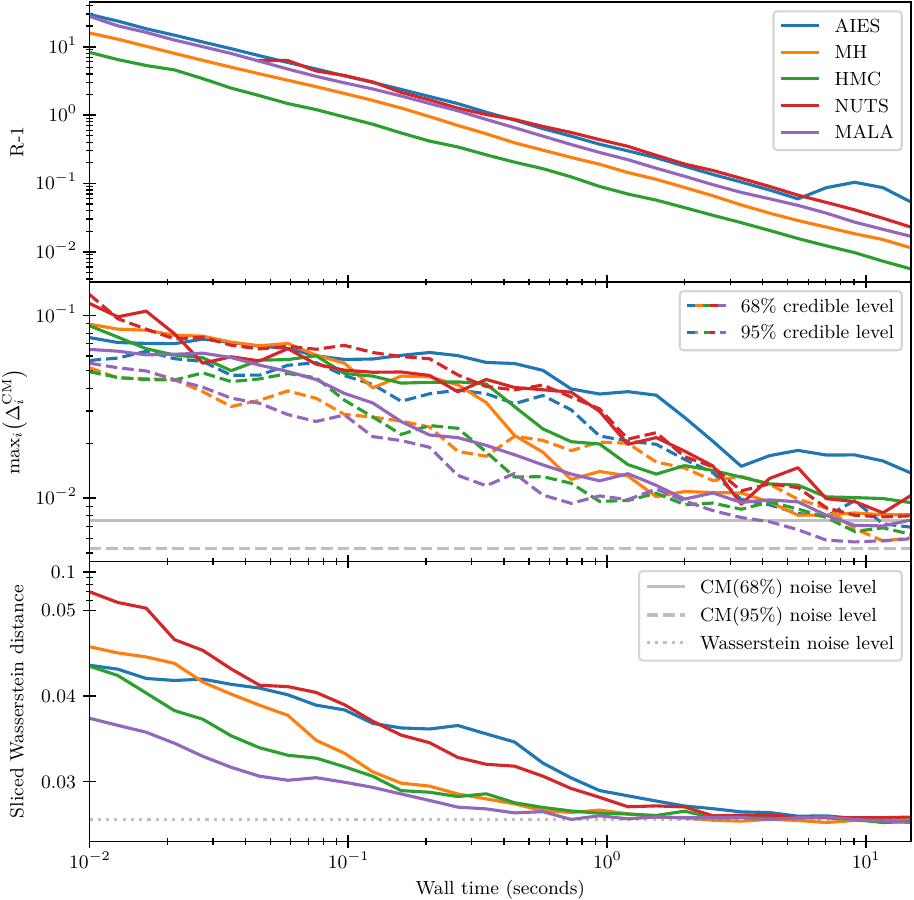}{fig:walltimes_lcdm}{All metrics as a function of wall time for every MCMC method using the $\Lambda$CDM emulator and 1,000 chains. Only the final samplings are depicted and the noise levels obtained from independent reference runs are depicted for the credible metric and the Wasserstein distance. $R-1$ for NUTS can not be computed before around 0.05 seconds of sampling due to the slow sampling speed.}

Figure~\ref{fig:samples_lcdm} likewise shows the convergence for all the MCMC methods but for all three metrics as a function of samples per chain and only showing the results for the samplings with 1,000 chains. Also shown are the expected noise levels for the credible metric and the sliced Wasserstein distance estimated from a second independent reference run. The general trend is that NUTS converges fastest (in terms of samples per chain) followed by HMC, MALA, and lastly MH and AIES. The gradient-based samplers, unsurprisingly, thus seem to be most efficient in exploring the parameter space in the least amount of steps. Notably, NUTS (and to some extent also HMC) seems to settle on a good covariance matrix already after a single burn-in iteration, whereas MALA and MH benefit from having all three iterations. Especially in the credible metric and the sliced Wasserstein distance, we see significant drops in the metric value after each burn-in iteration. One might think that increasing the number of burn-in iterations would result in a better initialisation of MH, MALA, and AIES, allowing them to match the level of NUTS and HMC in the beginning of the final sampling, but this is not the case. We have verified the covariance matrices obtained by each method (except for AIES that does not use the covariance matrix) after the burn-in period is very close to optimal, and initialising the methods with a covariance matrix computed from the reference run results in roughly the same initial metric values.

Since likelihood evaluations are no longer the dominant computational cost, wall time provides a more practically relevant comparison between samplers than convergence measured solely in terms of samples or likelihood evaluations. When taking the sampling speeds reported in table~\ref{tab:speeds} into account, the conclusion changes. Figure~\ref{fig:walltimes_lcdm} shows the same metrics but as a function of wall time instead (and without the burn-in phase). In terms of wall time, MALA now seems to be the fastest converging, closely followed by HMC, MH, and lastly, NUTS and AIES. Due to the complexity of NUTS and the resulting much slower speed, its convergence is very similar to that of AIES which was the slowest in terms of samples per chain. In fact, all the methods seem to converge very similarly in terms of wall time, so the most significant difference between running AIES or NUTS for 10 seconds is the amount of points sampled. The incredible sampling speeds of AIES and regular MH make them a viable alternative to more advanced approaches at the cost of having a significantly larger sample size due to their high autocorrelation lengths.

\subsection{Sterile neutrino results}

When sampling using the sterile neutrino extension emulator, we again initialise with three burn-in iterations (except for AIES having only 1) and sample 1,000 chains simultaneously. Due to this likelihood having two additional and highly non-Gaussian dimensions, the sampling and covariance estimation is more difficult. To accommodate this, the burn-in period has been increased in order to obtain good covariance matrices. In particular, the convergence of MALA is very sensitive to how well the covariance is estimated during burn-in for this specific likelihood. Using the same burn-in settings as for $\Lambda$CDM yields great results for all the other methods, but it was insufficient for MALA. We have nonetheless extended the burn-in periods of all other samplers to three iterations of 3,000 steps (of which the first 1,000 steps are used for step size adaptation and afterwards discarded) except for AIES still having only a single burn-in iteration but now of 20,000 steps.

\addfigure{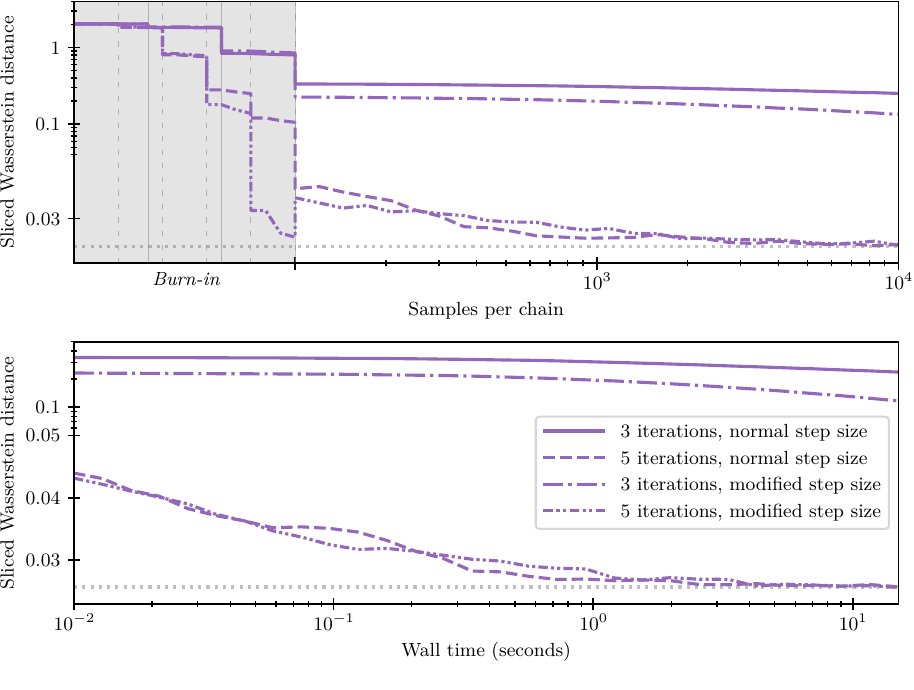}{fig:mala}{Sliced Wasserstein distance as a function of samples per chain (top) and wall time (bottom) for different burn-in configurations of MALA sampling with the sterile neutrino extension emulator. The dashed (solid) vertical lines in the top figure depicts 5 (3) burn-in iterations. Results are shown for the default step size configuration, where all dimensions (in whitened space) are sampled with the same step size, as well as for a modified step size where the sterile neutrino extension parameters are sampled with a 10 times larger step size.}

\addfigure{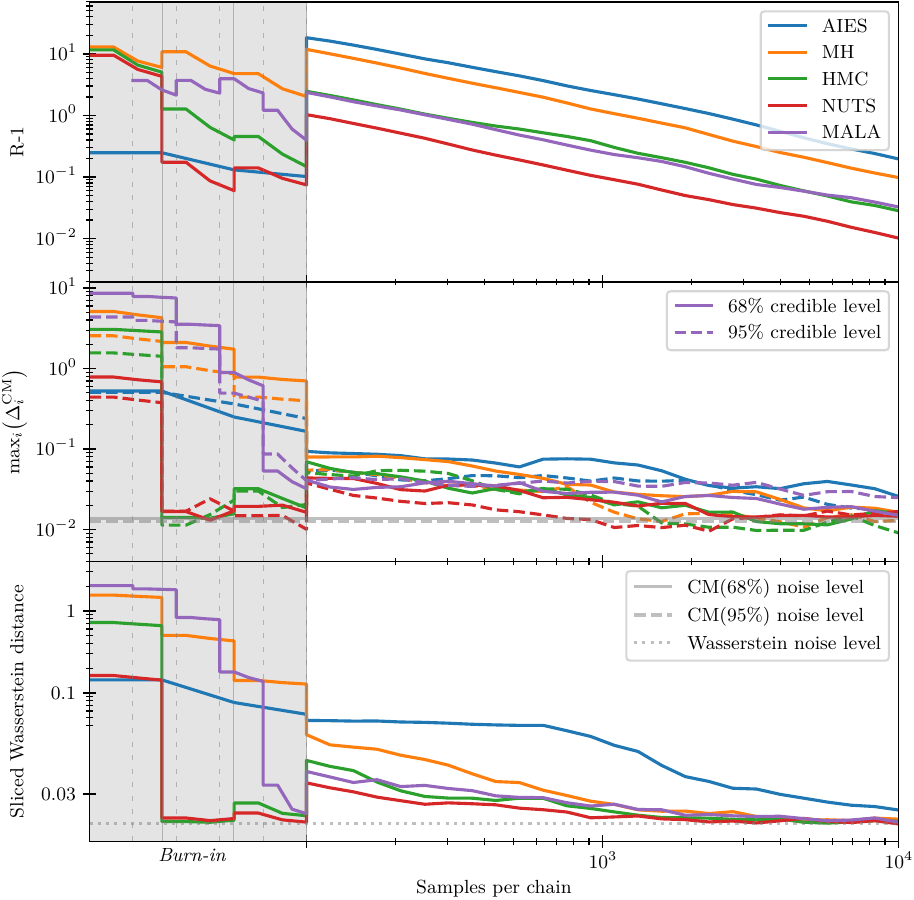}{fig:samples_sterile}{All metrics as a function of samples per chain for every MCMC method using the sterile neutrino extension emulator. The grey region refers to the burn-in period consisting of three updates of the covariance matrix before the final sampling, except for AIES which only has a single longer burn-in iteration. Noise levels obtained from independent reference runs are also depicted for the credible metric and the sliced Wasserstein distance. The vertical dashed lines depicts the 5 burn-in iterations of MALA necessary to obtain a good covariance matrix for this model.}

\addfigure{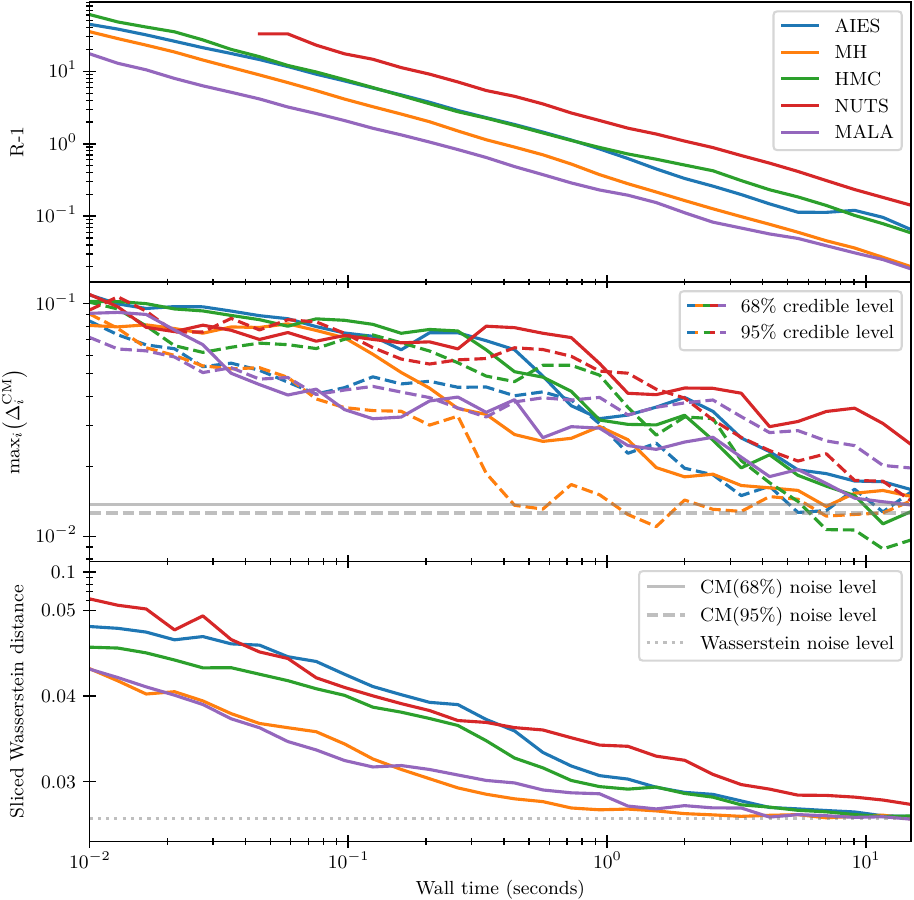}{fig:walltimes_sterile}{All metrics as a function of wall time for every MCMC method using the sterile neutrino extension emulator. Only the final samplings are depicted and the noise levels obtained from independent reference runs are depicted for the credible metric and the Wasserstein distance. $R-1$ for NUTS can not be computed before around 0.05 seconds of sampling due to the slow sampling speed.}

For MALA we have tested multiple burn-in configurations, and aside from increasing the number of burn-in iterations and the number of burn-in steps within each iteration, we can attempt to accommodate the sampling problems that arise due to the two extension parameters. The initial step size (in whitened space) is by default the same for all parameters, but due to the large plateaus of the extension parameters, it would be reasonable to increase the step size for these two dimensions alone\footnote{Aggressively increasing the step size for all parameters leads to suboptimal sampling and might result in very low acceptance rates for this likelihood.}. We have varied the number of burn-in iterations (3 or 5) as well as modified the initial step sizes of the extension parameters (increased by a factor of 10 compared to the rest of the parameters) while keeping the number of burn-in steps in each burn-in iteration fixed to 6,000 steps (of which the first 1,000 steps are used for step size adaptation and afterwards discarded). Figure~\ref{fig:mala} depicts the sliced Wasserstein distances for the four different configurations; 3 burn-in iterations and default step sizes, 5 burn-in iterations and default step sizes, 3 burn-in iterations and increased step sizes for the extension parameters, and 5 burn-in iterations and increased step sizes for the extension parameters. The number of burn-in iterations matter much more than modifying the step size and based on tests of varying the number of burn-in steps when a fixed number (3) of burn-in iterations, a smaller number of iterations cannot be made to work by simply increasing the number of steps within them. It is well known that MALA does not sample optimally for likelihoods like this where the gradient is not informative everywhere and the geometry varies much across the parameter space~\cite{pmlr-v238-biron-lattes24a}. Despite modifying the step sizes not having a major impact on the convergence, we include the results of the test with 5 burn-in iterations and a modified step size for the extension parameters in the following figures comparing with the results of the other MCMC methods.

Figure~\ref{fig:samples_sterile} shows the three metrics as a function of samples per chain. The burn-in periods are depicted here as well, and the 5 burn-in iterations of the MALA sampling are indicated by vertical dashed lines. The first iteration of MALA did not result in a sample with non-zero variance for all parameters, so $R-1$ could not be computed for that iteration. When allowing MALA to initialise with a good covariance matrix, we see roughly the same convergences as for $\Lambda$CDM. NUTS converges the fastest in terms of samples per chain and also this time, it seems to be well initialised even after a single burn-in iteration. The same is true for HMC even though it performs slightly worse compared to NUTS.

Turning to Figure~\ref{fig:walltimes_sterile}, it is again very clear that the slow sampling speed of NUTS completely negates the efficiency of its advanced algorithm in terms of wall time. Also for this likelihood it converges slower than the other methods, but does so, however, with fewer points. Due to their very high sampling speeds, MH and MALA converge very fast, but in turn, collects many more samples with larger autocorrelation length. In terms of wall time AIES converges similarly to HMC for this likelihood, but AIES will also collect many more samples in order to match HMC's convergence.

\section{Conclusion}\label{sec:conclusion}

The increasing use of neural-network emulators for cosmological inference changes some of the practical considerations involved in posterior sampling. In conventional analyses, where likelihood evaluations are computationally expensive, sampling efficiency is often assessed primarily in terms of the number of likelihood evaluations required to reach convergence. In the emulator setting considered here, likelihood evaluations become sufficiently inexpensive that performance in terms of wall time and hardware utilisation also become important factors. We have compared five MCMC methods (MH, MALA, HMC, NUTS, and AIES) using differentiable likelihood emulators implemented in \tf{}. By combining graph execution, JIT compilation, and GPU acceleration, all methods achieve sampling speeds that are substantially higher than those typically encountered in traditional cosmological inference pipelines.

When convergence is considered as a function of samples per chain, the gradient-based samplers perform best. NUTS generally requires the fewest samples to reach convergence, followed by HMC and MALA. This is consistent with the use of gradient information to guide exploration of parameter space and reduce random-walk behaviour. However, these gains in sample efficiency are partially offset by the additional computational cost of the more sophisticated algorithms. When convergence is instead considered as a function of wall time, the differences between the methods become considerably smaller. In this comparison, MALA often provides a favourable balance between computational cost and sampling efficiency, while the high throughput of MH allows it to remain competitive despite its comparatively poor sample efficiency. NUTS remains the most efficient method in terms of samples required for convergence, but its lower sampling speed reduces this advantage when measured in wall time.

A second observation is the usefulness of large-scale vectorisation. Running hundreds or thousands of chains simultaneously leads to faster convergence for all methods considered here, while the associated increase in wall time remains relatively modest on modern GPUs. This makes highly parallel sampling an attractive option when using emulator-based likelihoods.

Our results also illustrate some limitations of gradient-based approaches. While HMC and NUTS perform robustly for both likelihoods considered in this work, MALA becomes more sensitive to covariance estimation and burn-in for the sterile neutrino extension. The strongly non-Gaussian structure of this posterior appears to reduce the effectiveness of local gradient information and makes covariance adaptation more challenging. Nevertheless, once supplied with a sufficiently accurate covariance estimate or allowed enough burn-in iterations, MALA again performs competitively.

Overall, no single sampler is optimal in all situations. NUTS is the most sample-efficient method considered in this work, whereas MALA often performs favourably when convergence is measured in wall time. MH and AIES remain viable alternatives due to their simplicity and their ability to exploit vectorised likelihood evaluations efficiently. The choice of sampler therefore depends both on the properties of the posterior distribution and on the computational environment in which the sampling is performed. More generally, these results indicate that emulator-based cosmological inference shifts part of the focus from minimising likelihood evaluations to balancing sample efficiency, computational overhead, and parallel performance. In this setting, methods that would traditionally be regarded as inefficient can remain competitive when combined with highly parallel likelihood evaluations on modern hardware.

\section*{Acknowledgements}
We acknowledge computing resources from the Centre for Scientific Computing Aarhus (CSCAA). A.N. was supported by the Carlsberg Foundation, grant CF24-1944.


\bibliographystyle{utcaps}
\bibliography{sampling2026}

\providecommand{\href}[2]{#2}\begingroup\raggedright\begin{thebibliography}{10}

\bibitem{metropolis}
N.~Metropolis, A.~W. Rosenbluth, M.~N. Rosenbluth, A.~H. Teller, and E.~Teller,
  ``Equation of State Calculations by Fast Computing Machines,'' {\em Journal
  of Chemical Physics} {\bfseries 21} (1953) no.~6, 1087--1092.

\bibitem{hastings}
W.~K. Hastings, ``Monte Carlo sampling methods using Markov chains and their
  applications,'' \href{http://dx.doi.org/10.1093/biomet/57.1.97}{{\em
  Biometrika} {\bfseries 57} (1970) no.~1, 97--109},
  \href{http://arxiv.org/abs/http://biomet.oxfordjournals.org/cgi/reprint/57/1/97.pdf}{{\ttfamily
  http://biomet.oxfordjournals.org/cgi/reprint/57/1/97.pdf}}.
  \url{http://biomet.oxfordjournals.org/cgi/content/abstract/57/1/97}.

\bibitem{Zhou:2026eit}
Z.~Zhou, C.~Giovanetti, and H.~Liu, ``{ABCMB: A Python+JAX Package for the
  Cosmic Microwave Background Power Spectrum},''
  \href{http://arxiv.org/abs/2602.15104}{{\ttfamily arXiv:2602.15104
  [astro-ph.CO]}}.

\bibitem{Nygaard:2022wri}
A.~Nygaard, E.~B. Holm, S.~Hannestad, and T.~Tram, ``{CONNECT: a neural network
  based framework for emulating cosmological observables and cosmological
  parameter inference},''
  \href{http://dx.doi.org/10.1088/1475-7516/2023/05/025}{{\em JCAP} {\bfseries
  05} (2023)  025}, \href{http://arxiv.org/abs/2205.15726}{{\ttfamily
  arXiv:2205.15726 [astro-ph.IM]}}.

\bibitem{SpurioMancini:2021ppk}
A.~Spurio~Mancini, D.~Piras, J.~Alsing, B.~Joachimi, and M.~P. Hobson,
  ``{CosmoPower: emulating cosmological power spectra for accelerated Bayesian
  inference from next-generation surveys},''
  \href{http://dx.doi.org/10.1093/mnras/stac064}{{\em Mon. Not. Roy. Astron.
  Soc.} {\bfseries 511} (2022) no.~2, 1771--1788},
  \href{http://arxiv.org/abs/2106.03846}{{\ttfamily arXiv:2106.03846
  [astro-ph.CO]}}.

\bibitem{Gunther:2025xrq}
S.~G{\"u}nther, L.~Balkenhol, C.~Fidler, A.~R. Khalife, J.~Lesgourgues, M.~R.
  Mosbech, and R.~K. Sharma, ``{OL{\'E} {\textemdash} Online Learning Emulation
  in cosmology},'' \href{http://dx.doi.org/10.1088/1475-7516/2025/09/059}{{\em
  JCAP} {\bfseries 09} (2025)  059},
  \href{http://arxiv.org/abs/2503.13183}{{\ttfamily arXiv:2503.13183
  [astro-ph.CO]}}.

\bibitem{Bonici:2023xjk}
M.~Bonici, F.~Bianchini, and J.~Ruiz-Zapatero, ``{Capse.jl: efficient and
  auto-differentiable CMB power spectra emulation},''
  \href{http://arxiv.org/abs/2307.14339}{{\ttfamily arXiv:2307.14339
  [astro-ph.CO]}}.

\bibitem{Janken:2025wlq}
L.~Janken, S.~Hannestad, T.~Tram, and A.~Nygaard, ``{CLiENT: A new tool for
  emulating cosmological likelihoods using deep neural networks},''
  \href{http://arxiv.org/abs/2512.17509}{{\ttfamily arXiv:2512.17509
  [astro-ph.CO]}}.

\bibitem{Gammal:2022eob}
J.~E. Gammal, N.~Sch\"oneberg, J.~Torrado, and C.~Fidler, ``{Fast and robust
  Bayesian Inference using Gaussian Processes with GPry},''
  \href{http://arxiv.org/abs/2211.02045}{{\ttfamily arXiv:2211.02045
  [astro-ph.CO]}}.

\bibitem{mala}
G.~O. Roberts and J.~S. Rosenthal, ``Optimal Scaling of Discrete Approximations
  to Langevin Diffusions,''
  \href{http://dx.doi.org/10.1111/1467-9868.00123}{{\em Journal of the Royal
  Statistical Society Series B: Statistical Methodology} {\bfseries 60} (1998)
  no.~1, 255--268},
  \href{http://arxiv.org/abs/https://academic.oup.com/jrsssb/article-pdf/60/1/255/49589077/jrsssb\_60\_1\_255.pdf}{{\ttfamily
  https://academic.oup.com/jrsssb/article-pdf/60/1/255/49589077/jrsssb\_60\_1\_255.pdf}}.
  \url{https://doi.org/10.1111/1467-9868.00123}.

\bibitem{hmc}
S.~Duane, A.~D. Kennedy, B.~J. Pendleton, and D.~Roweth, ``{Hybrid Monte
  Carlo},'' \href{http://dx.doi.org/10.1016/0370-2693(87)91197-X}{{\em Phys.
  Lett. B} {\bfseries 195} (1987)  216--222}.

\bibitem{nuts}
M.~D. Hoffman and A.~Gelman, ``{The No-U-Turn Sampler: Adaptively Setting Path
  Lengths in Hamiltonian Monte Carlo},''
  \href{http://dx.doi.org/10.48550/arXiv.1111.4246}{{\em arXiv e-prints} (2011)
   }, \href{http://arxiv.org/abs/1111.4246}{{\ttfamily arXiv:1111.4246
  [stat.CO]}}.

\bibitem{10.2140/camcos.2010.5.65}
J.~Goodman and J.~Weare, ``{Ensemble samplers with affine invariance},''
  \href{http://dx.doi.org/10.2140/camcos.2010.5.65}{{\em Communications in
  Applied Mathematics and Computational Science} {\bfseries 5} (2010) no.~1, 65
  -- 80}. \url{https://doi.org/10.2140/camcos.2010.5.65}.

\bibitem{Gelman1996}
A.~Gelman, G.~O. Roberts, and W.~R. Gilks, ``Efficient Metropolis Jumping
  Rules,'' in {\em Bayesian Statistics 5}, J.~M. Bernardo, J.~O. Berger, A.~P.
  Dawid, and A.~F.~M. Smith, eds., pp.~599--608.
\newblock Oxford University Press, Oxford, 1996.

\bibitem{2023arXiv231207215R}
I.~Romero and M.~Ortiz, ``The energy-stepping Monte Carlo method: an exactly
  symmetry-preserving, a Hamiltonian Monte Carlo method with a 100\% acceptance
  ratio,'' \href{http://arxiv.org/abs/2312.07215}{{\ttfamily arXiv:2312.07215
  [math-ph]}}. \url{https://arxiv.org/abs/2312.07215}.

\bibitem{Radford2011}
R.~{Neal}, ``{MCMC Using Hamiltonian Dynamics},''.

\bibitem{2024arXiv240307495T}
J.~H. Tran and T.~S. Kleppe, ``Tuning diagonal scale matrices for HMC,''
  \href{http://arxiv.org/abs/2403.07495}{{\ttfamily arXiv:2403.07495
  [stat.CO]}}. \url{https://arxiv.org/abs/2403.07495}.

\bibitem{arnold1989mathematical}
V.~I. Arnold, \href{http://dx.doi.org/10.1007/978-1-4757-2063-1}{{\em
  Mathematical Methods of Classical Mechanics}}, vol.~60 of {\em Graduate Texts
  in Mathematics}.
\newblock Springer-Verlag, New York, 2~ed., 1989.
\newblock \url{https://doi.org/10.1007/978-1-4757-2063-1}.
\newblock Chapter 9: Canonical Formalism.

\bibitem{Goldstein2002}
H.~Goldstein, {\em Classical Mechanics}.
\newblock Addison-Wesley, 1980.

\bibitem{2017arXiv170102434B}
M.~Betancourt, ``{A Conceptual Introduction to Hamiltonian Monte Carlo},''
  \href{http://arxiv.org/abs/1701.02434}{{\ttfamily arXiv:1701.02434
  [stat.ME]}}.

\bibitem{apostol_calculus_vol2}
T.~M. Apostol, {\em Calculus, Volume 2: Multi-Variable Calculus and Linear
  Algebra with Applications to Differential Equations and Probability}.
\newblock Wiley, 2~ed., 1969.
\newblock Chapter 11: Multiple Intergrals.

\bibitem{2013PASP..125..306F}
D.~Foreman-Mackey, D.~W. Hogg, D.~Lang, and J.~Goodman, ``{emcee: The MCMC
  Hammer},'' \href{http://dx.doi.org/10.1086/670067}{{\em Publ. Astron. Soc.
  Pac.} {\bfseries 125} (2013)  306--312},
  \href{http://arxiv.org/abs/1202.3665}{{\ttfamily arXiv:1202.3665
  [astro-ph.IM]}}.

\bibitem{Planck:2018vyg}
{\bfseries Planck} Collaboration, N.~Aghanim {\em et al.}, ``{Planck 2018
  results. VI. Cosmological parameters},''
  \href{http://dx.doi.org/10.1051/0004-6361/201833910}{{\em Astron. Astrophys.}
  {\bfseries 641} (2020)  A6},
  \href{http://arxiv.org/abs/1807.06209}{{\ttfamily arXiv:1807.06209
  [astro-ph.CO]}}. [Erratum: Astron.Astrophys. 652, C4 (2021)].

\bibitem{Planck:2019nip}
{\bfseries Planck} Collaboration, N.~Aghanim {\em et al.}, ``{Planck 2018
  results. V. CMB power spectra and likelihoods},''
  \href{http://dx.doi.org/10.1051/0004-6361/201936386}{{\em Astron. Astrophys.}
  {\bfseries 641} (2020)  A5},
  \href{http://arxiv.org/abs/1907.12875}{{\ttfamily arXiv:1907.12875
  [astro-ph.CO]}}.

\bibitem{boss2016}
{\bfseries BOSS} Collaboration, S.~Alam {\em et al.}, ``{The clustering of
  galaxies in the completed SDSS-III Baryon Oscillation Spectroscopic Survey:
  cosmological analysis of the DR12 galaxy sample},''
  \href{http://dx.doi.org/10.1093/mnras/stx721}{{\em Mon. Not. Roy. Astron.
  Soc.} {\bfseries 470} (2017) no.~3, 2617--2652},
  \href{http://arxiv.org/abs/1607.03155}{{\ttfamily arXiv:1607.03155
  [astro-ph.CO]}}.

\bibitem{ross2014}
A.~J. Ross, L.~Samushia, C.~Howlett, W.~J. Percival, A.~Burden, and M.~Manera,
  ``{The clustering of the SDSS DR7 main Galaxy sample \textendash{} I. A 4 per
  cent distance measure at $z = 0.15$},''
  \href{http://dx.doi.org/10.1093/mnras/stv154}{{\em Mon. Not. Roy. Astron.
  Soc.} {\bfseries 449} (2015) no.~1, 835--847},
  \href{http://arxiv.org/abs/1409.3242}{{\ttfamily arXiv:1409.3242
  [astro-ph.CO]}}.

\bibitem{Beutler:2011hx}
F.~Beutler, C.~Blake, M.~Colless, D.~H. Jones, L.~Staveley-Smith, L.~Campbell,
  Q.~Parker, W.~Saunders, and F.~Watson, ``{The 6dF Galaxy Survey: Baryon
  Acoustic Oscillations and the Local Hubble Constant},''
  \href{http://dx.doi.org/10.1111/j.1365-2966.2011.19250.x}{{\em Mon. Not. Roy.
  Astron. Soc.} {\bfseries 416} (2011)  3017--3032},
  \href{http://arxiv.org/abs/1106.3366}{{\ttfamily arXiv:1106.3366
  [astro-ph.CO]}}.

\bibitem{10.1214/ss/1177011136}
A.~Gelman and D.~B. Rubin, ``{Inference from Iterative Simulation Using
  Multiple Sequences},'' \href{http://dx.doi.org/10.1214/ss/1177011136}{{\em
  Statistical Science} {\bfseries 7} (1992) no.~4, 457 -- 472}.
  \url{https://doi.org/10.1214/ss/1177011136}.

\bibitem{2019arXiv190200434K}
S.~Kolouri, K.~Nadjahi, U.~Simsekli, R.~Badeau, and G.~K. Rohde, ``Generalized
  Sliced Wasserstein Distances,''
  \href{http://arxiv.org/abs/1902.00434}{{\ttfamily arXiv:1902.00434 [cs.LG]}}.
  \url{https://arxiv.org/abs/1902.00434}.

\bibitem{cobaya}
J.~Torrado and A.~Lewis, ``{Cobaya: Code for Bayesian Analysis of hierarchical
  physical models},''
  \href{http://dx.doi.org/10.1088/1475-7516/2021/05/057}{{\em JCAP} {\bfseries
  05} (2021)  057}, \href{http://arxiv.org/abs/2005.05290}{{\ttfamily
  arXiv:2005.05290 [astro-ph.IM]}}.

\bibitem{brooks_gelman}
S.~P. Brooks and A.~Gelman, ``General Methods for Monitoring Convergence of
  Iterative Simulations,''
  \href{http://dx.doi.org/10.1080/10618600.1998.10474787}{{\em Journal of
  Computational and Graphical Statistics} {\bfseries 7} (1998) no.~4,
  434--455},
  \href{http://arxiv.org/abs/https://www.tandfonline.com/doi/pdf/10.1080/10618600.1998.10474787}{{\ttfamily
  https://www.tandfonline.com/doi/pdf/10.1080/10618600.1998.10474787}}.
  \url{https://www.tandfonline.com/doi/abs/10.1080/10618600.1998.10474787}.

\bibitem{pmlr-v238-biron-lattes24a}
M.~Biron-Lattes, N.~Surjanovic, S.~Syed, T.~Campbell, and A.~Bouchard-Cote,
  ``{autoMALA}: Locally adaptive {M}etropolis-adjusted {L}angevin algorithm,''
  in {\em Proceedings of The 27th International Conference on Artificial
  Intelligence and Statistics}, S.~Dasgupta, S.~Mandt, and Y.~Li, eds.,
  vol.~238 of {\em Proceedings of Machine Learning Research}, pp.~4600--4608.
\newblock PMLR, 02--04 May, 2024.
\newblock \url{https://proceedings.mlr.press/v238/biron-lattes24a.html}.

\end{thebibliography}\endgroup

\end{document}